\documentclass[twocolumn,floats,aps]{revtex4}
\usepackage{times,amsmath,amsfonts,psfrag,latexsym,graphics}

\begin{document}
\title{A Manganin Foil Sensor for Small Uniaxial Stress}
\author{M. K. Frampton, N. McLaughlin, Hu Jin, and R. J. Zieve$^*$}
\affiliation{ Physics Department, University of California, Davis, CA  95616, USA\\
${}^*$ Corresponding author}

\begin{abstract}
We describe a simple manganin foil resistance manometer for uniaxial stress
measurements. The manometer functions at low pressures and over a range
of temperatures.  In this design no temperature seasoning is necessary,
although the manometer must be prestressed to the upper end of the desired
pressure range. The prestress pressure cannot be increased arbitrarily;
irreversibility arising from shear stress limits its range. Attempting
larger pressures yields irreproducible resistance measurements.
\end{abstract}

\pacs{}
\maketitle

Manganin has been used and studied extensively as a pressure manometer for
decades, both for hydrostatic pressure and for shock waves. Its
resistivity is linear up to high pressures, described by a piezoresistance
coefficient $\frac{\Delta R}{RP}$. Near room temperature
the resistivity has little temperature-dependence, due to a nearby maximum
in the resistivity. Hence any temperature changes induced through
compression do not interfere with the reading of a resistance manometer
\cite{Adams}. Also, while the resistivity is temperature-dependent, at fixed
temperature it maintains its linearity in pressure \cite{Thompson}. 

Typical commercial manganin gauges use wire coils or grids manufactured from
very thin foil. Here we discuss a simple home-made manganin foil gauge for
uniaxial stress measurements. Our work involves lower pressures than usual
for manganin manometers, and we find that for our purposes various steps in
setting up the manometer become unnecessary. A main advantage is the ease of
making the gauge and adapting it to the geometry of the experiment.

\begin{figure}[]
\begin{center}
\hspace*{-.2in}\scalebox{0.5}{\includegraphics{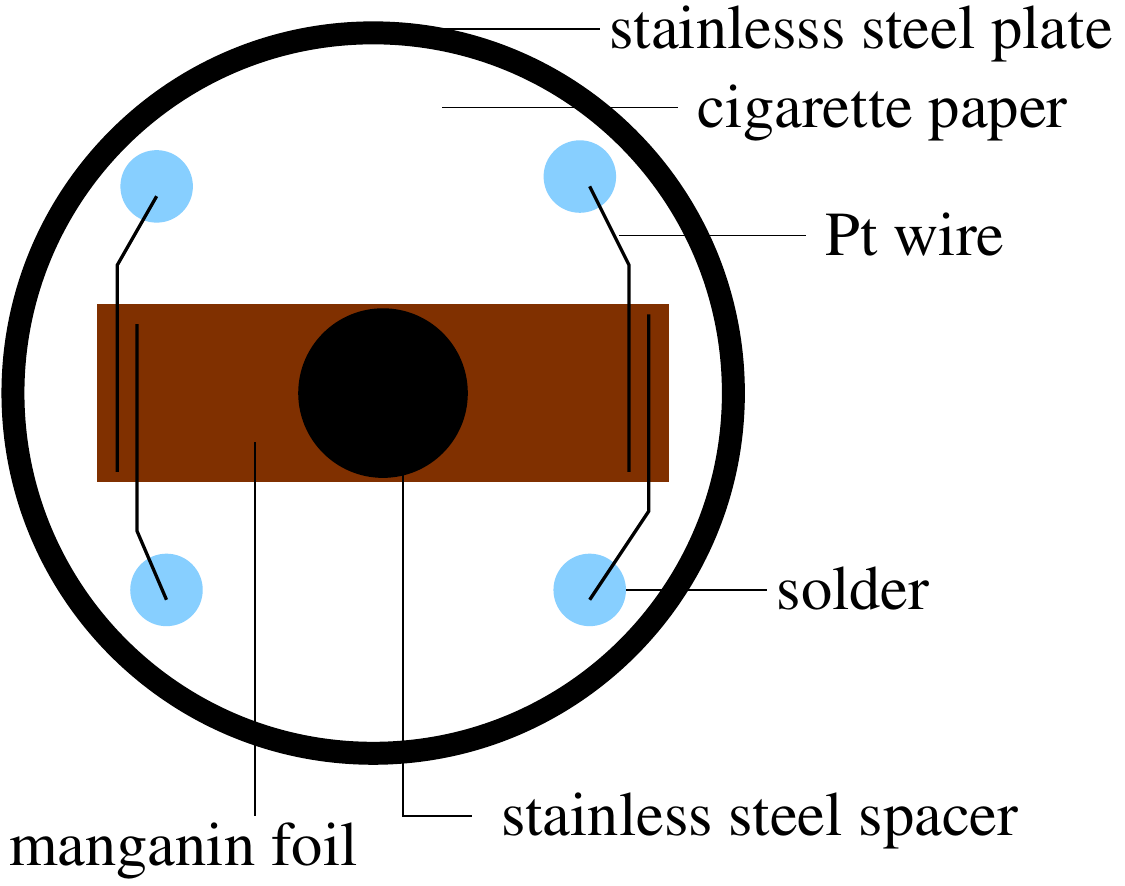}}\vspace{.3in}\\
\hspace*{.3in}\scalebox{0.5}{\includegraphics{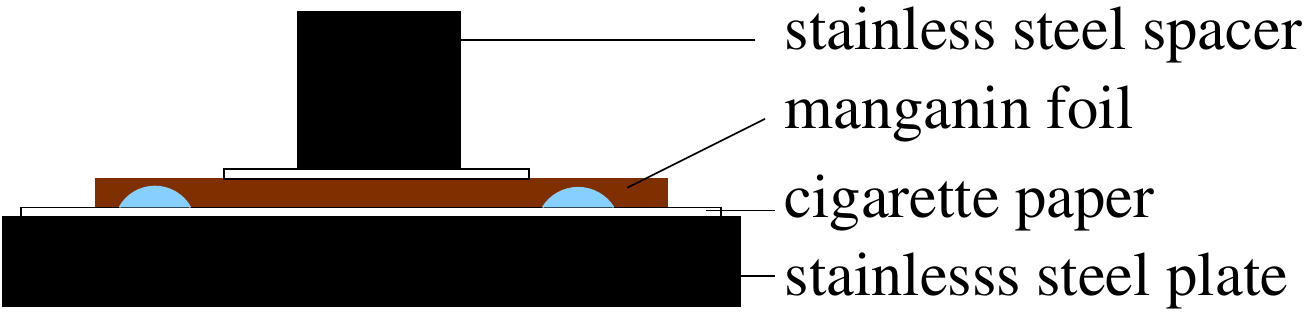}}
\caption{ 
Top and side views of the manganin manometer. Manganin foil is glued
to a stainless steel plate on one side and a stainless steel spacer on the other using thin layers of Stycast 1266 epoxy, across cigarette paper to ensure electrical isolation. Small platinum wires spot-welded
to the manganin connect to larger leads at solder pads. Stress is applied to the central region of the foil via the stainless steel spacer.}
\label{fig:setup}
\end{center}
\end{figure}

One such manometer, shown in Figure \ref{fig:setup}, is a 10 mm by 3 mm
piece of manganin foil, 0.5 mm thick. The foil is inserted into a
pressure column, with stress applied through a circular spacer to a
3 mm diameter region.  Platinum wires for resistance measurements are
spot-welded to the large top surface of the manganin foil, outside of
the stressed region, making these connections very simple.  Using a foil
larger than the pressure column also helps in assembling the column;
with a small piece it is more difficult to ascertain that the spacers
and manometer are properly positioned and aligned. The manganin
response is then used to determine the pressure on a sample located
elsewhere in the pressure column.

We use two different pressure setups for calibrating the manganin manometer
and confirming that its behavior is reproducible. In one the stress is
applied by tightening a screw; in the second it comes from pressurizing a
helium bellows \cite{Duke, Dix}. The latter setup is mounted on a
dilution refrigerator, but in the work here we use it from room
temperature down to liquid nitrogen temperature. Each apparatus has a
piezoelectric in the pressure column to measure the applied stress. After
calibration is complete we can transfer the manganin manometer to a
different screw-operated pressure cell, which has too small a diameter to
accommodate the piezoelectric sensors.

In previous experiments manganin pressure gauges often require
``seasoning" to give reproducible results. The treatment typically
includes both annealing to temperatures above and below room
temperature and also applying a higher pressure than will be reached
in the planned measurements. The high pressure may be applied multiple
times, and may be interspersed with the thermal cycling \cite{Adams,
Thompson, Zeto}.

In our uniaxial measurements, we find no effect from thermal
cycling. However, prestressing is crucial to obtaining acceptable
repeatability. We perform a single prestress at room temperature,
applying a pressure $P_{ps}$ of typically about 2.2 kbar and waiting
for a time from 2 to 40 minutes.  Lengthening the wait time has no
effect on subsequent measurements.  For sufficiently small $P_{ps}$,
the manometer's subsequent response is repeatable for pressures below
or even very slightly above $P_{ps}$. These pressures are extremely
low compared to previous work, but we find that seasoning at a higher
pressure does not expand the usable range of the manometer and can even
degrade its subsequent performance.

\begin{figure}[]
\begin{center}
\scalebox{.45}{\includegraphics{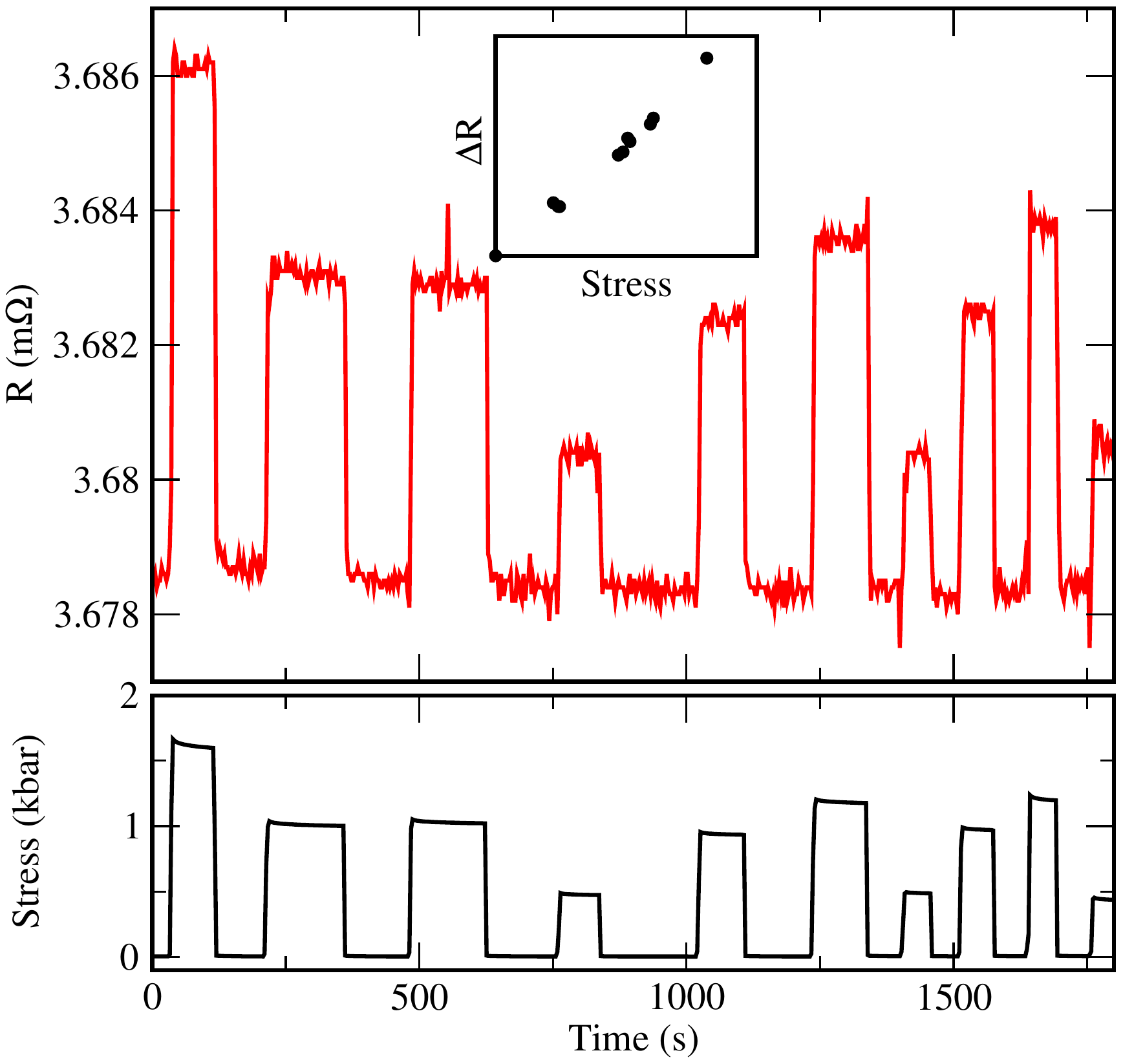}}
\caption{Upper: resistance of the manganin vs. time. Lower: applied stress, as measured with piezoelectric device. Inset: change in resistance as function of applied stress for these data, showing linear behavior.}
\label{fig:Robustness}
\end{center}
\end{figure}

After assembling the manometer and prestressing to a pressure $P_{ps}$, we
measure the resistance of the manganin foil as we apply pressures below
$P_{ps}$. We use several different pressures and apply each one
repeatedly, returning to zero pressure in between. This confirms the
reproducibility of the measurements and the linearity of $\Delta R$ with
applied pressure. One such trace appears in Figure \ref{fig:Robustness}.

To compare with previous work, we consider the pressure coefficient of
resistance $\frac{\Delta R}{RP}$, where $\Delta R=R-R_0$. For the manometer
with data shown here, the value is $7.8\times 10^{-4}$ kbar$^{-1}$. The
effect of pressure on $\Delta R/R$ should be independent of the manometer
geometry, as long as the pressure is applied to the entire manometer. Since
we apply pressure only to a portion of the current path for the resistance
measurement, we expect the pressure to have a reduced effect on $\Delta
R/R$. The separation between our inner contacts is a factor of two to three
larger than the diameter of the region under pressure. Since pressure
affects the resistivity along only a portion of the measured region, our
$\Delta R/R$ should be reduced by a factor of two to three, compared to the
value when pressure is applied to an entire manometer. This is consistent
with previous measurements for both uniaxial \cite{RPK, Rosenberg} and
hydrostatic \cite{Adams, Thompson} pressure, which find pressure
coefficients of resistance from $2\times 10^{-3}$ to $2.5\times 10^{-3}$
kbar$^{-1}$.

\begin{figure}[]
\begin{center}
\scalebox{.45}{\includegraphics{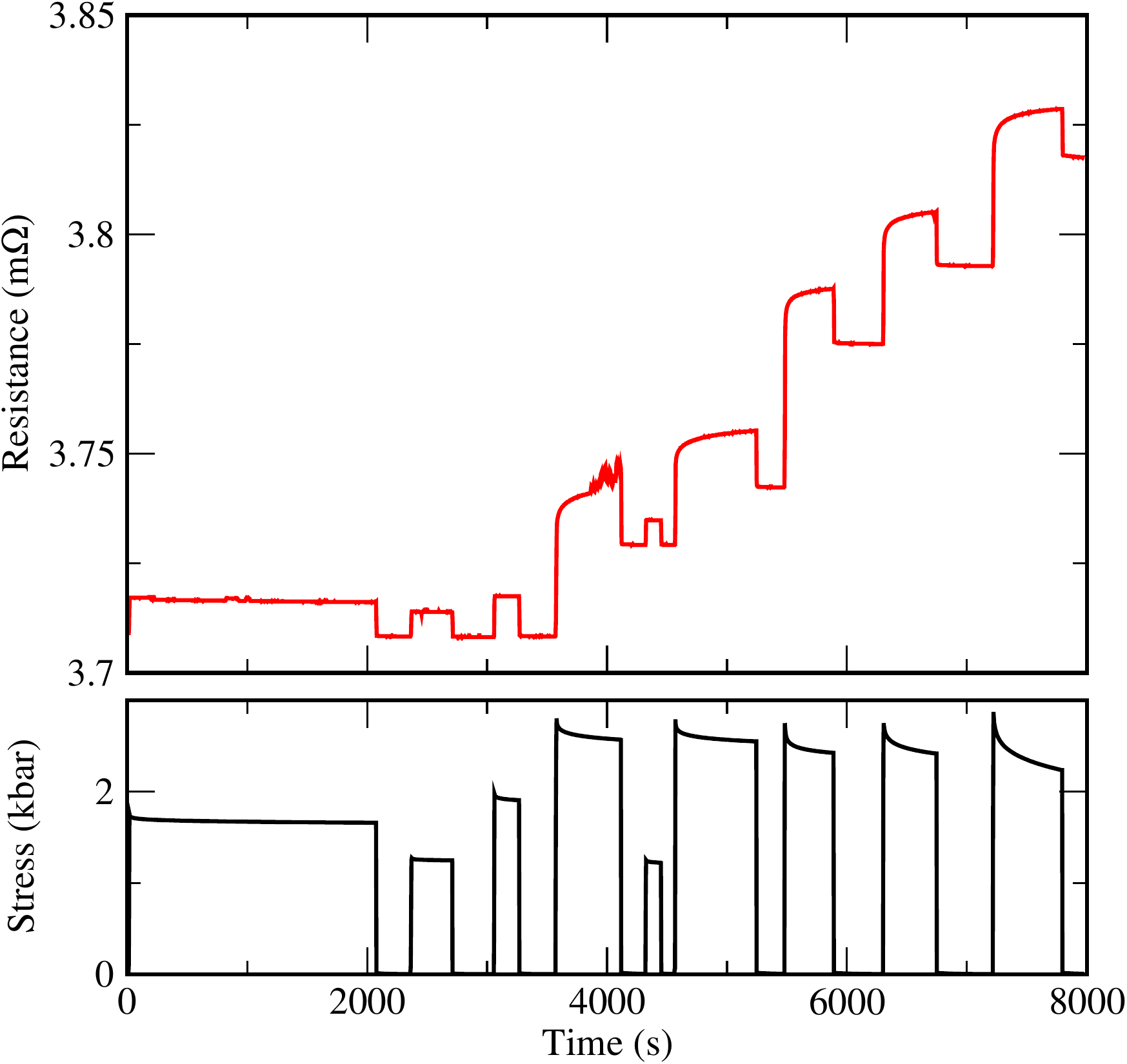}}
\caption{The top graph is the resistance of the manganin versus time,
which shows jumps according to when pressure is applied.  The bottom
graph shows the pressure applied to the manganin measured by the
piezometer versus time. The pressurizations at 3500s, 4500s, 5500s,
6300s, and 7200s all exceed the maximum prestress pressure of about
2kbar. In each case the resistance increases sharply and does not return
to its previous value after the pressure is released.}
\label{fig:OneTime}
\end{center}
\end{figure}

For applied pressures below the prestress pressure, we have performed
several dozen pressure cycles with no indication of a change in the pressure
coefficient of resistance or in the baseline resistance. This changes
dramatically if the applied pressure exceeds $P_{ps}$, or if the prestress
pressure exceeds a maximum effective pre-stress pressure $P_{max}$. For
$P_{ps}>P_{max}$, the behavior at pressures above $P_{max}$ but below
$P_{ps}$ is {\em not} reproducible and can lead to shifts in the floor
resistance. For this geometry, we find $P_{max}\approx 2$ kbar. Figure
\ref{fig:OneTime} exhibits this behavior, with the resistance rising sharply
each time a pressure above 2 kbar is applied. Removing the pressure shows
that the ambient-pressure resistance itself increases on each pressure
cycle. The pressure coefficient also decreases slightly after each
application of excess pressure. Furthermore, increasing the prestress
pressure does not eliminate the problem. Indeed, the five pressures above 2
kbar shown in Figure \ref{fig:OneTime} occur in decreasing order, but each
one nonetheless disrupts the manometer's prior behavior. 

Repeatable resistivity in manganin, linear with pressure, has been observed
up to 180 kbar under hydrostatic conditions \cite{Fujioka}. We attribute our
irreversibility at much lower pressure to the fact that we apply stress only
to a portion of the manganin foil.  Manganin is very sensitive to shear
stress, as indicated by variations in the pressure coefficient for nominally
identical setups once the pressure medium solidifies \cite{Samara}. The
circular edge of the our spacer defines a crossover between the pressurized
section of the foil and the remaining ambient-pressure region. The
cylindrical sheet lying directly below the circular edge is subjected to
significant shear stress, probably introducing defects in the foil. This is
consistent with the abrupt increase in resistance observed each time the
applied stress exceeds 2 kbar in Figure \ref{fig:OneTime}, as additional
defects are created. Upon repeated cycling we find no indication that resistance stabilizes, but the pressure coefficient decreases and the manometer becomes less useful. These observations are consistent with each pressure application creating new defects, primarily near the edge of the previous region of high defect density. As the defects begin to populate the portion of the manometer under the spacer, its pressure response changes.

To test the importance of the shear stress, we assembled smaller manganin
manometers with the spacers pressing on the entire foil. In this geometry we
found no maximum pressure limit for repeatability. Our apparatus could 
apply stress only up to about 4 kbar. Nonetheless, the small manometers easily
surpassed the $P_{max}$ observed for the larger devices. These observations
support previous evidence, mainly indirect, about the high sensitivity of
manganin to shear stress \cite{Adams, Samara}. Despite the pressure
limitation, for low pressures the larger manometer foil setup has a
significant advantage in ease of use. The smaller geometry requires
attaching the measurement leads to the thin sides of the foil, which is much
more challenging than placing them on the top surface. The small manometer
also creates difficulties in the proper alignment of the pressure column. If
the spacer piece slips to one side and acquires a slight angle, one edge of
the spacer can bypass the manometer entirely.

Another benefit of a manganin manometer is the ability to monitor the
stress directly at low temperature. In our screw-based pressure cells the
stress is always applied at room temperature but may change through
differential thermal contraction of the pressure cell materials. Empirically
the resistance is linear in pressure at each temperature. One previous group
reported further that the pressure coefficient is independent of
temperature \cite{Thompson}; others found that the pressure coefficient
increases with temperature \cite{Samara, Andersson}. In our setup the
pressure coefficient does have temperature-dependence. Nonetheless, our
resistance data as a function of both temperature and pressure fit the form
\[ R(p,T) = A + BT + CT^2 + Dp + EpT, \] which allows us to determine
pressure from measured temperature and resistance. The coefficients on
the right depend on the exact geometry, so each new manometer must be
calibrated in a trial run.

In addition to the manganin foil described above, we tested manganin
manometers constructed from wire and from thinner foil. Neither was as
successful. The wire manometer did not give reproducible results; this is
in fact consistent with previous claims that seasoning pressures for wire
manometers must exceed 4 kbar \cite{Adams, Fujioka}. The thinner foil,
0.2 mm, was sanded down from the thicker material. The main advantage
is a larger resistance, making measurements slightly easier. However,
the pressure coefficient itself did not change, and the thicker foil is easier to work with and faster to set up.

We present a simple construction of a manganin resistance manometer, for use
under low uniaxial stress. The design allows flexibility in the size and
shape of the manometer, such as reducing the active area to make the gauge
more sensitive when the pressure applied to a sample is very low. The
external portions can also easily be shaped to fit the pressure cell. The
setup requires only a single seasoning step of applying a pressure exceeding
the desired measurement pressures. It also enables pressure sensing at low
temperatures. The main limitation is that the stress on the gauge must
remain below about 2 kbar. We show that this unusually low threshold stems
from applied shear stress and confirm the sensitivity of manganin to shear.

{\bf ACKNOWLEDGEMENTS}

This work was supported by UC Davis through a Faculty Research Grant.
We thank L. Xiong for his work on temperature annealing of the foils.

\end{document}